\documentstyle[sprocl]{article}
\def\be{\begin{equation}}
\def\ee{\end{equation}}
\def\bea{\begin{eqnarray}}
\def\eea{\end{eqnarray}}

\begin{document}

\title{Acuasal Behavior in Quantum Electrodynamics}
\author{A. WIDOM [1], Y. N. Srivastava [1,2], E. Sassaroli [2,3]}
\address{1. Physics Department, Northeastern University, Boston, MA 02115, USA}
\address{2. Dipartimento di Fisica and sezione INFN \\ Universita' di 
Perugia, I-06123 Perugia, Italy}
\address{3. Laboratory for Nuclear Science and Department of Physics \\ 
Massachusetts Institute of Technology, Cambridge, MA 02139, USA}
\maketitle
\abstracts{
Acausal features of quantum electrodynamic processes are discussed. 
While these processes are not present for the {\it classical} electrodynamic 
theory, in the {\it quantum} electrodynamic theory, acausal processes are 
well known to exist. For example, any Feynman diagram with a ``loop'' in 
space-time describes a ``particle'' which may move forward in time 
or backward in time or in space-like directions. The engineering problems 
involved in experimentally testing such causality violations on a 
macroscopic scale are explored.
}

\section{Introduction}

The notion that signals can travel in both the forward and backward
directions in time (as well as space-like directions) has always appeared to
be an essential part of the unification of relativity and quantum mechanics.
The notion of an anti-particle as a physical particle traveling backwards in
time was at first conceived by St\"uckelberg \cite{1}, and later developed
by Feynman \cite{2} as an essential interpretation in a quantum field
theoretical frame work. To see what is involved, one may consider the free
photon propagator \cite{3} in the Feynman gauge, 
$$
D(x-y)=\int e^{iQ\cdot (x-y)}\left( {\frac{4\pi }{Q^2-i0^{+}}}\right) 
{\frac{d^4Q}{(2\pi )^4}}\ .\eqno(1.1) 
$$
By an elementary integration 
$$
D(x-y)=\frac i\pi \left( \frac 1{(x-y)^2+i0^{+}}\right) =\delta \left(
(x-y)^2\right) +\left( \frac i{\pi \left( x-y\right) ^2}\right) \ .\eqno(1.2)
$$
With $cT=x^0-y^0$ and $R=|{\bf x}-{\bf y}|$, the first term on the right
hand side of Eq.(1.2) yields one half the sum of the retarded and advanced
photon propagators 
$$
{\Re }eD(x-y)=
{\frac 12}\left( D_{retarded}(R,T)+D_{advanced}(R,T)\right) , \eqno(1.3) 
$$
$$
D_{retarded}(R,T)=\left( \frac{\delta (cT-R)}R\right) \ ,\ \
D_{advanced}(R,T)=\left( \frac{\delta (cT+R)}R\right) \ .\eqno(1.4) 
$$
Thus, the ``on light cone'' photon propagation may proceed in both the
forward (retarded) and backward (advanced) time directions. The last term on
the right hand side of Eq.(1.2) is non-vanishing in the time-like and
space-like directions, 
$$
{\Im }mD(x-y)=\left( \frac 1{\pi (R^2-c^2T^2)}\right) \ .\eqno(1.5) 
$$
Thus, ``off light cone'' space-like and time-like photon propagation is
feasible. Finally, when the photon propagator enters into a Feynman diagram
describing the amplitude for a process, all of the above propagation
directions, on and off the light cone, must be included to arrive at the
conventional accepted quantum electrodynamic results. Some physical
reasoning which helps in the understanding of why forward and backward in
time propagation enters into quantum field theory in an essential manner is
discussed in Sec.2. Space-like transmission of electromagnetic field
configurations are discussed in Sec.3. An electrical engineering
configuration for a transmission line device exhibiting space-like
transmission is discussed in Sec.4. Two photon causality violating
correlations are discussed in Sec.5. In the concluding Sec.6, some other
current views on causality will be briefly discussed.

\section{Measurable Quantum Fields are Non-Local in Time}

If $\phi (x)$ denotes a relativistic quantum field, then the usual procedure
for extracting the particle and/or anti-particle content from a field is to
decompose the field into positive and negative frequency parts as follows:
(i) Express $\phi (x)$ as a Fourier integral, using $x=({\bf r},ct)$, 
$$
\phi (x)=\int_{-\infty }^\infty 
\phi _\omega ({\bf r})e^{-i\omega t}d\omega . \eqno(2.1) 
$$
(ii) Divide the field into a positive frequency part (which destroys
particles and/or creates anti-particles) 
$$
\phi _{+}(x)=\int_0^\infty 
\phi _\omega ({\bf r})e^{-i\omega t}d\omega ,\eqno(2.2) 
$$
and a negative frequency part (which creates particles and/or destroys
anti-particles) 
$$
\phi _{-}(x)=\int_{-\infty }^0
\phi _\omega ({\bf r})e^{-i\omega t}d\omega , \eqno(2.3) 
$$
so that 
$$
\phi (x)=\phi _{+}(x)+\phi _{-}(x). \eqno(2.4) 
$$
The decomposition into positive and negative frequency parts can be written
using expressions in space-time \cite{4} by introducing a time-like velocity
four vector $v$, 
$$
v^\mu v_\mu =-c^2.\eqno(2.5) 
$$
Eqs.(2.2) and (2.3) then read (respectively) as 
$$
\phi _{+}(x)=\left( {\frac i{2\pi }}\right) \int_{-\infty }^\infty \phi
(x+v\tau )\left( \frac{d\tau }{\tau +i0^{+}}\right) ,\eqno(2.5) 
$$
and 
$$
\phi _{-}(x)=\left( {\frac i{2\pi }}\right) \int_{-\infty }^\infty \phi
(x-v\tau )\left( \frac{d\tau }{\tau +i0^{+}}\right) .\eqno(2.5) 
$$

To appreciate the experimental importance of this decomposition into
positive and negative frequency parts, it is sufficient to recall in quantum
optics that correlation functions of the form 
$$
{\cal G}_n({\bf r}_1,\lambda _1,t_1,...,{\bf r}_n,\lambda _n,t_n)= 
$$
$$
<E_{-}({\bf r}_1,\lambda _1,t_1)...E_{-}({\bf r}_n,\lambda _n,t_n)
E_{+}({\bf r}_n,\lambda _n,t_n)...E_{+}({\bf r}_1,\lambda _1,t_1)>, 
\eqno(2.6) 
$$
are thought\cite{5} to give a complete description of measurements for an
optical electric field 
$$
{\bf E}({\bf r},t)=\sum_{\lambda =x,y,z}E({\bf r},\lambda ,t){\bf e}_\lambda
,\eqno(2.7) 
$$
where 
$$
{\bf E}_{\pm }({\bf r},t)=\left( {\frac i{2\pi }}\right) 
\int_{-\infty}^\infty 
{\bf E}({\bf r},t\pm \tau )\left( \frac{d\tau }{\tau +i0^{+}} \right).
\eqno(2.8) 
$$
The important point is that to extract the photon production electric field 
${\bf E}_{-}({\bf r},t)$ and/or the photon detection electric field 
${\bf E}_{+}({\bf r},t)$ at time $t$, it is required that one know the 
physical electric field ${\bf E}({\bf r},t\pm \tau )$ both in the past 
and in the future of time $t$ \cite{6}$^,$\cite{7}. 
It is not sufficient to know the physical electric field ``now'' 
(at time $t$) to extract all experimental quantities. For example, 
the intensity of a light beam at space-time point 
$x=({\bf r},t)$ may be measured by 
$$
I({\bf r},t)=\left( \frac c{4\pi }\right) \sum_\lambda 
{\cal G}_1({\bf r},\lambda ,t)=\left( \frac c{4\pi }\right) 
<{\bf E}_{-}({\bf r},t){\bf \cdot E}_{+}({\bf r},t)>. 
\eqno(2.9) 
$$
Eqs.(2.6) and (2.8) require the electric field both in the past and in the
future of time $t$ of interest; Explicitly 
$$
I({\bf r},t)=\left( \frac c{16\pi ^3}\right) \int_{-\infty }^\infty
\int_{-\infty }^\infty 
{\frac{<{\bf E}({\bf r},t+s_1){\bf \cdot E}
({\bf r},t+s_2)>ds_1ds_2}{(s_1-i0^{+})(s_2+i0^{+})}}\ . 
\eqno(2.10) 
$$
From these considerations, it follows that the photon need not be located in
a spatial region in which there are presently electromagnetic fields. If
there will be in the future, or may have been in the past\cite{8},
electromagnetic fields in a given spatial region, then the future and past
conditions are sufficient for the photon to be located (with finite
probability) in a presently null field spatial region.

\section{Space-Like Transitions Between Field Configurations}

A difficulty that Einstein had with the conventional notions of quantum
electrodynamics was in part that the {\it Maxwell wave} associated with a
photon can be over here. Yet, the {\it photon} can be detected over there!
Sometimes, the region where the Maxwell wave for a photon is large turns out
to be a region without the photon (with finite probability). What troubled
Einstein also troubles us. All we can do is to show why Einstein's picture
of the photon and the associated Maxwell field is indeed what we presently
call conventional quantum electrodynamics. An initial photon field
configuration $|i>$ {\it over here} can be transported with superluminal
speed (space-like) to a final field configuration $|f>$ {\it over there}
with the finite quantum probability 
$$
P(i\to f)=|<f|i>|^2.\eqno(3.1) 
$$

If ${\bf e}({\bf r})$ and ${\bf b}({\bf r})$ denote and electromagnetic
field associated with a photon {\it at time zero}, then the mean energy of
the photon is given by 
$$
{\cal E}=\frac 1{8\pi }\int 
\left( |{\bf e}({\bf r})|^2+|{\bf b}({\bf r})|^2\right) . 
\eqno(3.2) 
$$
Employing the complex field 
$$
{\bf F}({\bf r})={\bf e}({\bf r})+i{\bf b}({\bf r}), 
\eqno(3.3) 
$$
and its Fourier transform 
$$
{\bf F}({\bf r})=\int {\bf \Psi }({\bf k})e^{i{\bf k\cdot r}}
\left( \frac{d^3k}{(2\pi )^3}\right), \eqno(3.4) 
$$
one finds the mean photon energy 
$$
{\cal E}={\frac 1{8\pi }}\int |{\bf F}({\bf r})|^2d^3r
={\frac 1{8\pi }}\int |
{\bf \Psi }({\bf k})|^2\left( \frac{d^3k}{(2\pi )^3}\right).
\eqno(3.5) 
$$
The photon wave function (Heisenberg at time zero) ${\bf \Psi }({\bf k})$ in
momentum space $({\bf p}=\hbar {\bf k})$ is here normalized as 
$$
{\frac 1{8\pi \hbar c}}\int |{\bf \Psi }({\bf k})|^2\left( \frac{d^3k}{(2\pi
)^3|{\bf k}|}\right) =1;\eqno(3.6) 
$$
e.g. $|{\bf \Psi }({\bf k})|^2(d^3k/|{\bf k}|)$ is proportional to the
probability of finding the photon with momentum in the Lorentz invariant
phase space element $(d^3k/|{\bf k}|)$. Note that the condition 
${\bf k\cdot }{\bf \Psi }({\bf k})=0$ is equivalent to the vacuum Maxwell 
equations ${\bf \nabla \cdot F}({\bf r})=0$. 
For two normalized photon wave functions, 
${\bf \Psi }_i({\bf k})$ 
(initial photon over here), and ${\bf \Psi }_f({\bf k})$
(final photon over there), {\it all at time zero}, the ``overlap'' or
transition amplitude is given by 
$$
<f|i>={\frac 1{8\pi \hbar c}}
\int {\bf \Psi }_f^{*}({\bf k}){\bf \cdot \Psi }_i({\bf k})
\left( \frac{d^3k}{(2\pi )^3|{\bf k}|}\right) . 
\eqno(3.7) 
$$
The notion of ``over here'' and ``over there'' has more to do with the
electromagnetic fields in space 
$$
{\bf \Psi }_{f,i}({\bf k})=\int {\bf F}_{f,i}({\bf r})
e^{-i{\bf k\cdot r}}d^3r, 
\eqno(3.8) 
$$
i.e. 
$$
<f|i>={\frac 1{16\pi ^3\hbar c}}\int \int \frac{{\bf F}_f^{*}({\bf r})
{\bf \cdot F}_i({\bf s})}{|{\bf r}-{\bf s}|^2}\ d^3rd^3s, 
\eqno(3.9) 
$$
where 
$$
{\bf F}_f^{*}({\bf r})={\bf e}_f({\bf r})-i{\bf b}^*_f({\bf r}),\ \ \ 
{\bf F}_i({\bf s})={\bf e}_i({\bf s})+i{\bf b}_i({\bf s}). \eqno(3.10) 
$$
The central result of this section follows from Eqs.(3.1), (3.9) and (3.10).
Let ${\bf e}_i({\bf r})$ and ${\bf b}_i({\bf r})$ denote respectively the
electric and magnetic fields associated with an initial photon (over here)
and let ${\bf e}_f({\bf r})$ and ${\bf b}_f({\bf r})$ denote respectively
the electric and magnetic fields associated with a final photon (over
there). The regions ``over here'' and ``over there'' mean that the initial
and final fields have compact support in non-overlapping spatial regions.
This situation is present at time zero. The transition probability
(space-like) obeys 
$$
P(i\to f)= 
$$
$$
{\frac 1{256\pi ^6\hbar ^2c^2}}\left| {\int \int }
\frac{\left( {\bf e}_f({\bf r})-i{\bf b}^*_f({\bf r})\right) 
{\bf \cdot }\left( {\bf e}_i({\bf s})+
i{\bf \cdot b}_i({\bf s})\right) }{|{\bf r}-{\bf s}|^2}{\ d^3rd^3s}\ 
\right|^2\ .\eqno(3.11) 
$$
Thus, with finite probability, the photon can go from over here to over
there (space-like) in no time at all. Note that the space like propagation
of photons in Eq.(1.5) enters in the space-like transition probability of
Eq.(3.11) in thinly disguised form. For zero times, such that $x=({\bf r},0)$
and $y=({\bf s},0)$, the imaginary part of the propagator, 
$\pi {\Im }mD(x-y)=|{\bf r}-{\bf s}|^{-2}$, provides the space-like 
transition probability kernel in the central Eq.(3.11). 
Material photon propagators exist that are more efficient 
(for space-like transitions) than the vacuum propagator.

\section{Transmission Lines}

Consider an electromagnetic transmission line along the $z$-axis \cite{9}.
The line voltage at point $z$ on the line is given by Faraday's law 
$$
v(z,t)=-{\frac 1c}\left( \frac{\partial \phi (z,t)}{\partial t}\right). 
\eqno(4.1) 
$$
If $\varepsilon $ denotes the line capacitance per unit length, and if $\mu $
denotes the line inductance per unit length, then the transmission line 
$(1+1)$-dimensional Lagrange density is given by 
$$
{\cal L}={\frac \varepsilon {2c^2}}\left( \frac{\partial \phi }{\partial t}
\right) ^2-{\frac 1{2\mu }}\left( \frac{\partial \phi }{\partial z}\right)
^2.\eqno(4.2) 
$$
The line charge density (per unit length) 
$$
\rho (z,t)=\varepsilon v(z,t)\eqno(4.3) 
$$
plays the role of a conjugate field to $\phi (x)$ with an equal time
commutation relation 
$$
[\rho (z,t),\phi (z^{\prime },t)]=i\hbar c\delta (z-z^{\prime }).\eqno(4.4) 
$$
The transmission line signal velocity 
$$
u={\frac c{\sqrt{(\varepsilon \mu )}}},\ \ \ \varepsilon \mu \ge 1,\eqno(4.5)
$$
enters into the transmission line wave equation 
$$
{\frac 1{u^2}}\left( \frac{\partial ^2\phi }{\partial t^2}\right) =\left( 
\frac{\partial ^2\phi }{\partial z^2}\right) .\eqno(4.6) 
$$
Electromagnetic transmission lines, (in classical theory) propagate signals
somewhat slower ($u\le c$) than vacuum light speed. In order to understand
the notion of a transmission line impedance, we first consider the notion of
a vacuum impedance. The vacuum Maxwell equations have the form 
$$
\partial _\mu F^{\mu \nu }=-R_{vac}J^\nu ,\eqno(4.7) 
$$
which defines the vacuum impedance $R_{vac}$. In terms of the electronic
charge $e$, the quantum electrodynamic coupling strength reads 
$$
\alpha =\left( \frac{e^2R_{vac}}{4\pi \hbar }\right) .\eqno(4.8) 
$$
Eqs.(4.7) and (4.8) hold true in any set of units. The vacuum impedance is
always defined in terms of light speed $c$. For example, in the Gaussian
units here employed, the vacuum has an impedance 
$$
R_{vac}=\left( \frac{4\pi }c\right) \simeq
419.169004390336362426121257964335\ (picosec/cm).\eqno(4.9) 
$$
In engineering SI units 
$$
R_{vac}={\frac 1{\epsilon _0c}}=\mu _0c\simeq
376.7303134617706554681984004203193\ (Ohms).\eqno(4.10) 
$$
(The value $R_{vac}=(1/c)=1.0000000000000000000000000000000000$ is employed
and easily remembered in high energy physics.) The transmission line
impedance (in Gaussian units) is defined as 
$$
R=\left( \frac 1{\varepsilon u}\right) =\left( \frac{\mu u}{c^2}\right)
=\left( \frac{R_{vac}}{4\pi }\right) \sqrt{\frac \mu \varepsilon }\ .
\eqno(4.11) 
$$
A typical laboratory cable (say connecting a computer to the outside world)
is a transmission line with $R\approx 50\ Ohms$, or in Gaussian units 
$R\approx 55.6\ (picosec/cm)$. 

Of interest here is the possibility in {\it quantum} electrodynamic theory
of sending a superluminal signal down a $50\ Ohm$ cable. For a cable of line
impedance $R$ and line velocity $u$, we use a $(1+1)$ dimensional vector
notation $x=(z,ut)$. The photon propagator for an infinite transmission line
is defined as 
$$
D(x-y)={\frac i{\hbar c}}<0|\phi (x)\phi (y)|0>_{+},\eqno(4.12) 
$$
where the $+$ indicates time ordering. The transmission line propagator
obeys an equation of motion, 
$$
-\partial _\mu \partial ^\mu D(x-y)=cR\delta (x-y),\eqno(4.13) 
$$
which may be solved by the Fourier transformation 
$$
D(x-y)=\int e^{iQ\cdot (x-y)}\left( \frac{cR}{Q^2-i0^{+}}\right) 
\frac{d^2Q}{(2\pi )^2}\ .\eqno(4.14) 
$$

Strictly speaking, the integral in Eq.(4.14) does not exist, which leads the
mathematician to conclude that $(1+1)$-dimensional massless quantum field
theories do not exist. In turn, this causes the philosopher to ponder about
whether $50\ Ohm$ cables exist. The situation is similar to $(1+1)$ 
-dimensional strings. A mathematician cannot embed a quantum $(1+1)$ 
-dimensional string in a $(3+1)$-dimensional world. This leads the
philosopher to question whether a musical violin with strings can exist in a
world with three-dimensional Euclidean geometry. Infinite $50\ Ohm$ cables
do {\it not} exist, and the finite length of the physical cable serves as a
cut-off to the quantum electrodynamic theory. One can employ (for
mathematical simplicity) an ``infinite cable'' model of a long (but finite)
cable at the expense of introducing a large regulator length $\Lambda $ into
the intermediate stages of the computation. Such an engineering
approximation is quite all right if the regulator length does not enter into
the final physical answer. 

The propagator 
$$
D(x-y)=\left( \frac{icR}{4\pi }\right) ln\left( \frac{\Lambda ^2}
{(x-y)^2+i0^{+}}\right) \eqno(4.15) 
$$
is a solution to Eq.(4.13). The equal time correlation (Weightman)) function
for the cable of impedance $R$, 
$$
W(z-z^{\prime })=\frac 1{\hbar c}<0|\phi (z,0)\phi (z^{\prime },0)|0>, 
\eqno(4.16) 
$$
is then evaluated as 
$$
W(z)=\left( \frac{cR}{2\pi }\right) ln\left| \frac \Lambda z\right| ,\ \ \
(|z|<<\Lambda ).\eqno(4.17) 
$$
Eq.(4.17) may be safely employed for physical problems in which only the
difference 
$$
W(z_1)-W(z_2)=\left( \frac{cR}{2\pi }\right) ln\left| \frac{z_2}{z_1}\right|
.\eqno(4.18) 
$$
enters into the final answer since the regulator length $\Lambda $ is not
present in such differences. Now, let us consider placing a charge density
per unit length $\lambda (z)$ onto the transmission line. If $|0>$ denotes
the transmission line ground state, then the state with a charge density 
$\lambda (z)$ on the line may be defined as 
$$
\Big|\lambda \Big>=\exp \left( \frac i{\hbar c}\int \lambda (z)\phi
(z)dz\right) \Big|0\Big>.\eqno(4.19) 
$$
If we place an initial charge density $\lambda _i(z)$ on the cable ``over
here'' and we wish to compute the amplitude for finding a final charge
density $\lambda _f(z)$ on the cable ``over there'', then the transition
amplitude for such superluminal transport of charge density is given by 
$$
\left\langle \lambda _f\Big|
\lambda _i\right\rangle =\left\langle 0\right| \exp \left( \frac i{\hbar
c}\int \left( \lambda _i(z)-\lambda _f(z)\right) \phi (z)dz\right) \left|
0\right\rangle =\exp (-S_{fi}/\hbar ).\eqno(4.20) 
$$
The Euclidean (space-like) action $S_{fi}$ entering into Eq.(4.20) can be
evaluated since the ground state wave function $|0>$ implies a Gaussian
probability distribution for $\phi (z)$; i.e. 
$$
S_{fi}={\frac 1{2\hbar c^2}}\int \int \left( \lambda _i(z)-\lambda
_f(z)\right) \left( \lambda _i(z^{\prime })-\lambda _f(z^{\prime })\right)
<0|\phi (z)\phi (z^{\prime })|0>dzdz^{\prime }.\eqno(4.21) 
$$
Eqs.(4.16) and (4.22) may be written as 
$$
S_{fi}={\frac 1{2c}}\int \int \left( \lambda _i(z)-\lambda _f(z)\right)
\left( \lambda _i(z^{\prime })-\lambda _f(z^{\prime })\right) W(z-z^{\prime
})dzdz^{\prime }.\eqno(4.22) 
$$
For the case of an initial charge density over here being displaced a
distance $b$ to over there; i.e. 
$$
\lambda _i(z)=\lambda (z),\ \ \ \lambda _f(z)=\lambda (z+b),\eqno(4.23) 
$$
the Euclidean action is given by 
$$
S[\lambda ]={\frac 1{2c}}\int \int \lambda (z)\lambda (z^{\prime })\left(
2W(z-z^{\prime })-W(z-z^{\prime }-b)-W(z-z^{\prime }+b)\right) dzdz^{\prime
}.\eqno(4.24) 
$$
For the (large distance $b$) space-like transport of a voltage $\tilde
v(z)=\lambda (z)/\epsilon $, Eqs.(4.18) and (4.24) imply the Euclidean
action 
$$
S[\lambda ]={\frac R{4\pi }}\int \int \lambda (z)\lambda (z^{\prime
})ln\left( \left| \frac b{z-z^{\prime }}\right| ^2-1\right) dzdz^{\prime },\
\ \ \ b>>|z-z^{\prime }|.\eqno(4.25) 
$$
The superluminal transition probability is then 
$$
P(b)=\left| <\lambda _f|\lambda _i>\right| ^2=exp(-2S[\lambda ]/\hbar ). 
\eqno(4.28) 
$$
If the number of electrons $N$ which take part in the superluminal
transition is defined as 
$$
Ne=\int \lambda (z)dz,\eqno(4.29) 
$$
and if $a$ denotes the spread in space of the initial charge density, then
the probability for a superluminal transition of the voltage $\tilde
v(z)=\lambda (z)/\varepsilon $ through the distance $b$, obeys the decay law 
$$
P(b)\approx \left( \frac ab\right) ^\beta ,\ \ \ (a<<b).\eqno(4.30) 
$$
The decay exponent is given by 
$$
\beta =\left( \frac{e^2R}{\pi \hbar }\right) N^2=4\alpha \left( \frac
R{R_{vac}}\right) N^2,\eqno(4.31) 
$$
where Eq.(4.8) has been employed. The central result of this section is that
the exponent 
$$
\beta \approx 0.004\times N^2\ \ \ if\ \ \ R\approx 50\ Ohms.\eqno(4.32) 
$$
If the number of electrons involved in a signal obeys $N>>1$ then $\beta >>1$
and the probability of superluminal transport falls very rapidly with
spatial distance. In quantum electrodynamic theory, this large value of 
$\beta $ explains why it is not very easy to obtain superluminal transport 
on a $50\ Ohm$ cable. On the other hand, for only a few electrons 
$\beta \le 1$, which looks interesting except that one would have to 
settle for a very weak signal. We here leave the engineering considerations 
at this point.

\section{Photon Correlations}

Often, during the course of quantum photon experiments, one looks for
correlations between the number of photons detected in a two different
photon counters, say detectors $1$ and $2$. The photon coincidence
correlation function is then 
$$
C_{12}=<N_1N_2>,\eqno(5.1) 
$$
where $N_i$, for $i=1,2$, are the number operators for the photons in the
detectors. For a typical case, such as a Hanbury-Brown-Twiss experiment, the
correlation function $C_{12}$ depends upon the positions of the two
detectors. A typical application of the correlation $C_{12}$ measuring
``coincidence photon counts'' is made in astrophysics where the two photons
come from one and/or the other of two possible stars. It is difficult to
resolve the two stars from straight forward single detector intensity
measurements. Feynman\cite{10} has analyzed such two detector experiments as
follows: Since two photons are detected in a coincidence count (one photon
in each of the two counters), there are four physical possibilities: (i)
Both photons come from star $1$ with amplitude $a_{11}$. (ii) Both photons
come from star $2$ with amplitude $a_{22}$. (iii) The photon from star $1$
went to detector $1$ and the photon from star $2$ went to detector $2$ with
amplitude $a_{12}$. (iv) The photon from star $1$ went to detector $2$ while
the photon from star $2$ went tom detector $1$ with amplitude $a_{21}$ The
probability of the coincidence count is then 
$$
P(coincidence)=|a_{11}|^2+|a_{22}|^2+|a_{12}+a_{21}|^2,\eqno(5.2) 
$$
illustrating the quantum rules that one adds probabilities for
distinguishable events and adds amplitudes for indistinguishable events
(before taking the absolute value squared). Quantum interference between the
exchange amplitudes (the cross terms when absolute value squaring the last
term on the right hand side of Eq.(5.2)) allows for the resolution of the
positions of ``two relatively incoherent stars''.

Feynman's method of ``counting or listing possibilities on your fingers''
and then calculating quantum probabilities works equally well for more high
technology coherent photon sources. For example, suppose that one has two
photon coherent sources which are guaranteed to fire off {\it exactly two
photons} at a time. Such sources exist in laboratories within present
quantum optics technology. Let us further suppose that there are four and
only four possibilities for each two photon firing event \cite{11}: (i) Two
photons both go to detector $1$, i.e. $N_1=2,\ N_2=0$. (ii) Two photons both
go to detector $2$, i.e. $N_1=0,\ N_2=2$. (iii) Photon $1$ goes to detector 
$1$ and photon $2$ goes to detector $2$, i.e. $N_1=1,\ N_2=1$. (iv) Photon $1$
goes to detector $2$ and photon $2$ goes to detector $1$, i.e. $N_1=1,\
N_2=1 $. The last two possibilities have amplitude interference leading to a
non-trivial (and measured) correlation function 
$$
C({\bf r}_1,{\bf r}_2)=<2|N_1N_2|2>\eqno(5.3) 
$$
where ${\bf r}_i$, for $i=1,2$, are the detector position. We use the Dirac
notation $|2>$ to remind the reader that for the above four possibilities we
have 
$$
(N_1+N_2)|2>=2|2>.\eqno(5.4) 
$$
One easily proves the following \medskip 

\noindent 
{\bf Spooky Theorem:} The counting statistics at counter $1$ depends on how
one sets the position of counter $2$.

\noindent 
{\bf Spooky Proof:} From Eqs.(5.3) and (5.4) it follows that 
$$
2<2|N_1|2>-<2|N_1^2|2>=C({\bf r}_1,{\bf r}_2).\eqno(5.5) 
$$
What makes the theorem spooky is that the counting statistics at counter $1$
depends on the position of counter $2$ via counter $2$ events that are
possibly ``space-like'' or possibly ``in the future'' of counter $1$ events.

\section{Conclusion}

We have discussed above several examples of why it appears that conventional
quantum electrodynamics allows for interactions to proceed forward and
backward in time as well as space-like in direction. While the results are
conventional, the consequences are abhorrent to many. Einstein concluded
(from what he regarded as the clairvoyant nature of quantum mechanics) that
there are pieces of the puzzle missing in our present picture; i.e. quantum
mechanics is presently an incomplete view. Those less revolutionary than was
Einstein \cite{12}, prefer to think that these terms in quantum mechanics
that look like causality violations are present only in the mathematics but
not in the laboratory. One might hear that space-like photon propagation is
merely virtual. This closing of the eyes, ears, and mind may satisfy some
workers who do not like to think about what should be unthinkable; i.e. that
the future can effect the past and so forth. One sometimes hears a timid
statement that light can go faster than light speed ... but not really!
Other methods for a theoretical approach to acausality imply that causality
violations are certainly possible. But let us choose a system\cite{13} well
out of the reach of normal research laboratories. For example, if only we
could make a worm hole at an outrageous density of $10^{100}$ ({\em who
cares whose units?}) which is not possible, then we would have a real time
machine.

We hope that our discussion of more realistic examples (such as a $50\ Ohm$
transmission line) may more quickly give rise to a serious engineering view
of the matter.

\end{document}